\documentclass[jr]{sjp}
\usepackage{graphicx} 
\usepackage{amsmath,amssymb}
\usepackage{float}
\floatstyle{boxed} 
\usepackage{dsfont}
\usepackage{tikz}
\usepackage{caption}
\usepackage{standalone}
\begin{document}
\mark{{Real Space Renormalization Group for One-Dimensional Ising Chains}{Shraddha Singh and Vijay A. Singh}}
\title{Real Space Renormalization Group for One-Dimensional Ising Chains}
\author[]{Shraddha Singh$^1$ and Vijay Singh$^2$$^\ast$\footnote{$^\ast$physics.sutra@gmail.com}}
\address[]{Integrated BS-MS Student, 2014-19, Department of Physics, UM-DAE, Centre for Excellence in Basic Sciences, Mumbai University, Kalina, Santa-Cruz East, Mumbai 400 098, India.\\
$^2$Raja Ramanna Fellow, Department of Physics, UM-DAE, Centre for Excellence in Basic Sciences, Mumbai University, Kalina, Santa-ruz East, Mumbai 400 098, India.}  
\begin{abstract}
We apply the real space Renormalisation Group (RNG) technique to a variety of one-dimensional Ising chains. We begin by recapitulating the work of Nauenberg for an ordered Ising chain, namely the decimation approach. We extend this work to certain non-trivial situation namely, the Alternate Ising Chain and Fibonacci Ising chain. Our approach is pedagogical and accessible to undergraduate students who have had a first course in statistical mechanics.
\end{abstract}
\maketitle
\section{Introduction}
There is a deep and useful connection between Statistical Mechanics and Quantum Field Theory. Kenneth Wilson appreciated this connection and applied the renormalization ideas to statistical mechanics\cite{1}. Application of these techniques to both classical and quantum many body problems have seen success. However, RNG calculations are often very complex and the approximations made are sometimes obscure. Often, one has to resort to extensive numerical calculations.
 
\par The present work is written in the spirit of conveying some essential ideas of RNG to a beginner and applying this approach to more complicated Ising chains. We present some pedagogical examples of a form of real space RNG termed Decimation. This technique was introduced by Michael Nauenberg in the context of the one-dimensional Ising model.  Unfortunately, this attractive piece of work \cite{2} is marked by several typographical errors. We present Nauenberg's work in a simplified (and hopefully error-free) fashion. We extend it to related Hamiltonians such as the  Alternate Ising model and Fibonacci chain Ising model.

\par The RNG strategy can be symbolically stated as follows. It transforms the Hamiltonian, e.g. $H^{'}=\mathds{R}(H)$. Next, one iterates it, $H^{''}=\mathds{R}(H')$ until one obtains a fixed point Hamiltonian, $H^{*}=\mathds{R}(H^{*}).$ The flow towards the fixed point Hamiltonian and the Hamiltonian $H^{*}$ itself yields insight into the physical properties of the system. Wilson suggested such a procedure and was able to elicit the critical properties of the 2D and the 3D Ising model and a famous quantum system namely, the Kondo problem\cite{3}. 

In Sec. 2, we recapitulate the work of Nauenberg and describe how decimation is carried out for the one-dimensional Ising model. In Sec. 3, we extend this approach to alternate Ising model where the coupling is alternate like in a binary alloy. In Sec. 4, we discuss the Fibonacci Ising chain. Sec. 5 constitutes the conclusion.

\section {One Dimensional Ising Model}
We start with the familiar one-dimensional Ising model for N spins, $S_i=\pm1$, $i=1,2...N$, with nearest neighbour coupling constant J, see Fig.(1).
\newline
\begin{figure}[h]
\centering
\begin{tikzpicture}
\node (1) [circle, draw, fill=lightgray] at (0,0) {S1};
\node (2) [circle, draw, fill=lightgray] at (2,0) {S2};
\draw(1)--(2) node[pos=0.5,sloped,above] {$J$};
\node (3) [circle, draw, fill=lightgray] at (4,0) {S3};
\draw(2)--(3) node[pos=0.5,sloped,above] {$J$};
\node (4) [circle, draw, fill=lightgray] at (6,0) {S4};
\draw(3)--(4) node[pos=0.5,sloped,above] {$J$};
\node (5) [circle, draw, fill=lightgray] at (8,0) {S5};
\draw(4)--(5) node[pos=0.5,sloped,above] {$J$};

\end{tikzpicture}
\caption{One Dimensional Ising Spin model}
\end{figure}
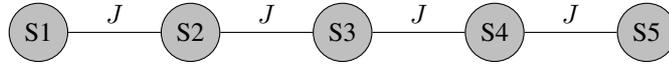
The Hamiltonian $H_N$ for this model is written as,
\begin{eqnarray}
H_N&=&-\frac{J}{kT}\sum_{i=1}^{N}S_iS_{i+1},
\end{eqnarray}
where $S_{N+1}=S_1$, J is the nearest neighbiour exchange coupling, T is the temperature, and k is the Boltzmann constant. We divide the coupling constant by kT for the sake of convenience in further derivations. One can consider the dimensionless Hamiltonian $H_N$, without  loss of generality,
\begin{eqnarray}
H_N(K)&=&-K\sum_{i=1}^{N}S_iS_{i+1}\quad\quad(\frac{J}{kT}=K) 
\end{eqnarray}
 Note that $K>0$ implies ferromagnetism.
\subsection{Decimation}
Let $\mathds{P}$ be the transfer matrix such that $\mathds{P}(i,i+1)=\text{exp}(KS_iS_{i+1})$.
Thus, the canonical partition function $Z_N$ is given by,
\begin{eqnarray} 
Z_N&=&\sum_{s_1,s_2,s_3...}exp(-H_N(K))=\sum \mathds{P}(S_1S_2)\mathds{P}(S_2S_3)\mathds{P}(S_3S_4)... \\
\mathds{P}&=&
  \left[ {\begin{array}{cc}
   e^K & e^{-K} \\
   e^{-K} & e^K \\
  \end{array} }\right]
\end{eqnarray}
As the elements of the matrix depend on the product $S_iS_{i+1}$ which is same for all i, we can write
\begin{eqnarray} 
Z_N=\sum_{s_1,s_2,s_3...}(\mathds{P(K)}^N)=Tr (\mathds{P(K)}^N)
\end{eqnarray}
Now, instead of computing the usual partition sum as shown above, we consider only the partial sum of $exp[-H_N(K)]$ over all possible values of  even spins, $S_i=\pm1, i=2,4,....$ and for even N we obtain a scaled partition function $exp[-H_N(K')]$, (see Fig. (2)).
\newline
\begin{figure}[h]
\centering
\begin{tikzpicture}
\node (1) [circle, draw, fill=lightgray] at (0,2) {S1};
\node (2) [circle, draw, fill=lightgray] at (2,2) {S2};
\draw(1)--(2) node[pos=0.5,sloped,above] {$K$};
\node (3) [circle, draw, fill=lightgray] at (4,2) {S3};
\draw(2)--(3) node[pos=0.5,sloped,above] {$K$};
\node (4) [circle, draw, fill=lightgray] at (6,2) {S4};
\draw(3)--(4) node[pos=0.5,sloped,above] {$K$};
\node (5) [circle, draw, fill=lightgray] at (8,2) {S5};
\draw(4)--(5) node[pos=0.5,sloped,above] {$K$};

\node (1) [circle, draw, fill=lightgray] at (0,0) {S1};
\node (2) [circle, draw, fill=lightgray] at (4,0) {S3};
\draw(1)--(2) node[pos=0.5,sloped,above] {$K'$};
\node (3) [circle, draw, fill=lightgray] at (8,0) {S5};
\draw(2)--(3) node[pos=0.5,sloped,above] {$K'$};

\draw[->](4,1.5)--(4,0.5);

\end{tikzpicture}
\caption{Decimation}
\end{figure}
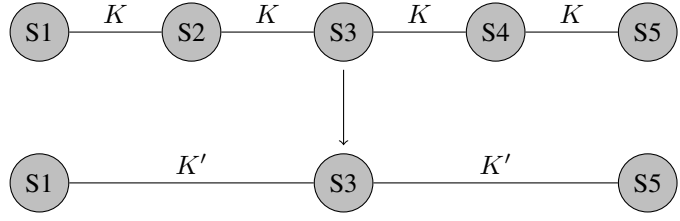
\newline
The Hamiltonian becomes, 
\begin{eqnarray}
\sum_{[s_2s_4..s_N]}exp[-H_N(K)]=\mathds{P}_{S_1S_3}^2\mathds{P}_{S_3S_5}^2....\mathds{P}_{S_{N-1}S_1}^2
\end{eqnarray}
\par The idea behind this partial summation is to find a renormalization transformation K$\,\to\, K^{'}$ such that,
\begin{eqnarray}
 \mathds{P}^2(K)&=&exp[2g(K)]\mathds{P}(K^{'})\\
\sum exp(-H_N(K))&=&Tr \mathds{P(K)}^N= Tr [\mathds{P(K)}^2]^{\frac{N}{2}},
\end{eqnarray}
 where g(K) is a scalar function of K. Then $K^{'}$ can be interpreted as an effective Ising coupling constant for the remaining  odd spins $S_i, i=1,3,5....N-1$ and Eq. (8)  may (formally) be written as.
\begin{eqnarray}
exp(-H_N(K))&=&\mathds{P(K)}^N\\
&=&[exp(2g(K))\mathds{P(K')}]^{N/2}\\
&=&exp(Ng(K))[(\mathds{P}(K'))]^{N/2}\\
&=&exp(Ng(K))exp(-H_{\frac{N}{2}}(K'))
\end{eqnarray}
Thus, the resulting equation becomes,
\begin{eqnarray}
\sum_{[s_1s_2..s_N]}exp[-H_N(K)]&=&\sum_{[s_1s_3..s_N]}exp[-H_{N/2}(K^{'})+Ng(K)]
\end{eqnarray}
To make the procedure clear we discuss the case of 3 spins,
\begin{eqnarray}
e^{K'S_1S_3}*e^{2g(K)}&=&\sum_{S_2=+1}^{-1}e^{KS_1S_2}*e^{KS_2S_3} \\
&=&e^{K(S_1+S_3)}+e^{-K(S_1+S_3)}
\end{eqnarray}
If $S_1=S_3=+1$
\begin{eqnarray}
e^{K'}*e^{2g(K)}=e^{2K}+e^{-2K}
\end{eqnarray}
If $S_1=-S_3=+1$
\begin{eqnarray}
e^{-K'}*e^{2g(K)}=2
\end{eqnarray}
Using Eq.(17) we obtain g(K) 
\begin{eqnarray}
g(K)=\frac{1}{2}K^{'}+\frac{1}{2}\text{ln2}
\end{eqnarray}
Next using Eq. (16) we obtain K'
\begin{eqnarray}
K^{'}=\frac{1}{2}\text{ln\{cosh(2K)}\}
\end{eqnarray}
Thus $K'$ is related to the original coupling K by a non-linear transformation. We denote this as $K'=f(K)$. Near the fixed point $K = K^* +\epsilon$,
\begin{eqnarray}
K'&=&f(K^*+\epsilon )\\
K^*+\epsilon '&=&f(K^*)+\epsilon f'(K^*)
\end{eqnarray}
As $K^*=f(K^*)$ near a fixed point K* we have 
\begin{eqnarray}
\epsilon '=\epsilon f'(K^*)
\end{eqnarray}
which is a linear transformation that resembles 
\begin{eqnarray}
\epsilon '=\lambda\epsilon
\end{eqnarray}
where $\lambda=\text{tanh}(2K^*)$. 
\par There are two solutions for the equation $K^*=\text{ln}\{\text{cosh}(2K^*)\}/2$, which are known as fixed points, $K^*=0$ and $K^*=\infty$ with $\lambda=0$ and $\lambda=1$ respectively.
For phase transition $\lambda$ must be greater than unity. This proves the well established result that there is no phase transition for 1D Ising spin model. There is another way to see this. After applying the renormalization transformation n times, the mapping $K^{n-1}\,\to\, K^{(n)}$ can be obtained from the recurrence relation, 
\begin{eqnarray}
K^{(n)}=\frac{1}{2}\text{ ln} {\text{ \{cosh(2K}^{(n-1)})\}}
\end{eqnarray}
where $K^{(0)}=K$. \\ 
\begin{eqnarray}
\text{Let }\zeta&=&\text{tanh(K)}\\
\mbox{therefore } K'&=& \frac{1}{2}\text{ln}(\frac{1+\zeta^2}{1-\zeta^2})\\
\text{Hence, } \zeta'=\text{tanh(K')}&=&\text{tanh}\{ \frac{1}{2}\text{ln}(\frac{1+\zeta^2}{1-\zeta^2})\}\\
&=&\dfrac{\text{exp}\Big( \dfrac{1}{2}\text{ln}\Big(\dfrac{1+\zeta^2}{1-\zeta^2}\Big)\Big)-\text{exp}\Big(-\Big( \dfrac{1}{2}\text{ln}\Big(\dfrac{1+\zeta^2}{1-\zeta^2}\Big)\Big)\Big)}{\text{exp}\Big( \dfrac{1}{2}\text{ln}\Big(\dfrac{1+\zeta^2}{1-\zeta^2}\Big)\Big)+\text{exp}\Big(-\Big( \dfrac{1}{2}\text{ln}(\dfrac{1+\zeta^2}{1-\zeta^2}\Big)\Big)\Big)}\\
&=&\dfrac{\sqrt{\dfrac{1+\zeta^2}{1-\zeta^2}}-\sqrt{\dfrac{1-\zeta^2}{1+\zeta^2}}}{\sqrt{\dfrac{1+\zeta^2}{1-\zeta^2}}+\sqrt{\dfrac{1-\zeta^2}{1+\zeta^2}}}\\
&=&\frac{(1+\zeta^2)-(1-\zeta^2)}{(1+\zeta^2)+(1-\zeta^2)}
\end{eqnarray}
Thus,
\begin{eqnarray}
\zeta '&=&\zeta^{2}\\
\text{tanh(K')}&=&\text{tanh(K)}^2
\end{eqnarray}
Since tanh(K) $<$ 1, tanh(K$^{n}$) tends to zero as $n \longrightarrow \infty$. This suggests that the effective coupling gets weaker with each decimation and we are left with a non-itneracting system which will show no phase transition.  We next discuss the more complex (unequal $J_i$) one-dimensional Ising models.
\section{Alternate Ising model}
\par Here the $K_i's$ are arrranged in the manner shown in Fig.(3).
\newline
\begin{figure}[h]
\centering
\begin{tikzpicture}
\node (1) [circle, draw, fill=lightgray] at (0,0) {S1};
\node (2) [circle, draw, fill=lightgray] at (6,0) {S4};
\draw(1)--(2) node[pos=0.5,sloped,above] {$K_1'$};
\node (3) [circle, draw, fill=lightgray] at (12,0) {S7};
\draw(2)--(3) node[pos=0.5,sloped,above] {$K_2'$};

\node (1) [circle, draw, fill=lightgray] at (0,2) {S1};
\node (2) [circle, draw, fill=lightgray] at (2,2) {S2};
\draw(1)--(2) node[pos=0.5,sloped,above] {$K_1$};
\node (3) [circle, draw, fill=lightgray] at (4,2) {S3};
\draw(2)--(3) node[pos=0.5,sloped,above] {$K_2$};
\node (4) [circle, draw, fill=lightgray] at (6,2) {S4};
\draw(3)--(4) node[pos=0.5,sloped,above] {$K_1$};
\node (5) [circle, draw, fill=lightgray] at (8,2) {S5};
\draw(4)--(5) node[pos=0.5,sloped,above] {$K_2$};
\node (6) [circle, draw, fill=lightgray] at (10,2) {S6};
\draw(5)--(6) node[pos=0.5,sloped,above] {$K_1$};
\node (7) [circle, draw, fill=lightgray] at (12,2) {S7};
\draw(6)--(7) node[pos=0.5,sloped,above] {$K_2$};

\draw[->](6,1.5)--(6,0.5);

\end{tikzpicture}
\caption{Similar Decimation for Alternate Ising Model}
\end{figure}
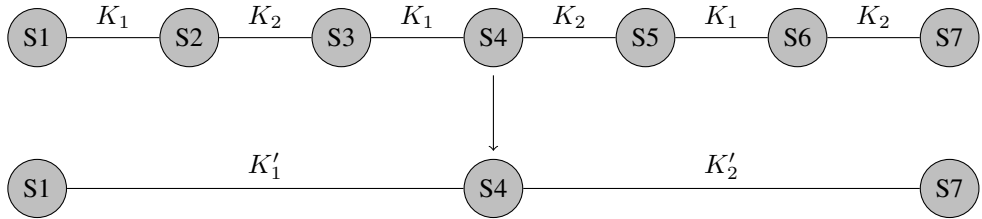
\\In order to adopt a similar decimation procedure we need to consider four spins at a time. This is illustrated in Fig.(3). Using this procedure,
\begin{eqnarray}
e^{K_1'S_1S_4}*e^{g_1}&=&\sum_{S_2,S_3}e^{K_1S_1S_2}*e^{K_2S_2S_3}*e^{K_1S_3S_4}\\
&=&\sum_{S_3}e^{K_1S_3S_4}*2\text{cosh}(K_1S_1+K_2S_3)\\
&=&e^{K_1S_4}*2cosh(K_1S_1+K_2)+e^{-K_1S_4}*2\text{cosh}(K_1S_1-K_2)
\end{eqnarray}
Like in the previous section, we consider $S_1=S_4=+1$ to obtain ,
\begin{eqnarray}
e^{K_1'}*e^{g_1}=e^{K_1}*2\text{cosh}(K_1+K_2)+e^{-K_1}*2\text{cosh}(K_1-K_2)
\end{eqnarray}
and $S_1=-S_4=+1$ to obtain,
\begin{eqnarray}
e^{-K_1'}*e^{g_1}=e^{-K_1}*2\text{cosh}(K_1+K_2)+e^{K_1}*2\text{cosh}(K_1-K_2)
\end{eqnarray}
Using Eqs.(36) and (37), and cosh(x)=cosh(-x) we obtain,
\begin{eqnarray}
e^{2K_1'}=\frac{e^{K_1}\text{cosh}(K_1+K_2)+e^{-K_1}\text{cosh}(K_1-K_2)}{e^{K_1}\text{cosh}(K_1-K_2)+e^{-K_1}\text{cosh}(K_1+K_2)}\end{eqnarray}
employing  the addition properties of cosh function, we obtain, 
\begin{eqnarray}
\text{tanh}(K_1')=\text{tanh}^2(K_1)\text{tanh}(K_2)
\end{eqnarray}
One similarly obtains,
\begin{eqnarray}
\text{tanh}(K_2')=\text{tanh}^2(K_2)\text{tanh}(K_1)
\end{eqnarray}
The fixed points are,$\{K_1^*,K_2^*\}=\{0,0\}$ or $\{\infty,\infty\}$.
\par Note that we need to block spins in a judicious way. If we block them in a non similar fashion, i.e. if the new lattice is not alternate (see Fig.(4)), then,
\newline
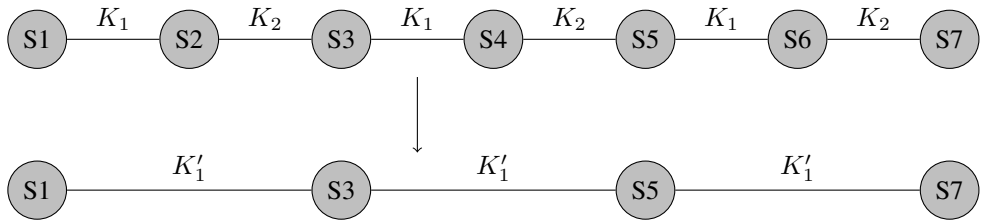
\begin{figure}[h]
\centering
\begin{tikzpicture}
\node (1) [circle, draw, fill=lightgray] at (0,0) {S1};
\node (2) [circle, draw, fill=lightgray] at (4,0) {S3};
\draw(1)--(2) node[pos=0.5,sloped,above] {$K_1'$};
\node (3) [circle, draw, fill=lightgray] at (8,0) {S5};
\draw(2)--(3) node[pos=0.5,sloped,above] {$K_1'$};
\node (4) [circle, draw, fill=lightgray] at (12,0) {S7};
\draw(3)--(4) node[pos=0.5,sloped,above] {$K_1'$};

\node (1) [circle, draw, fill=lightgray] at (0,2) {S1};
\node (2) [circle, draw, fill=lightgray] at (2,2) {S2};
\draw(1)--(2) node[pos=0.5,sloped,above] {$K_1$};
\node (3) [circle, draw, fill=lightgray] at (4,2) {S3};
\draw(2)--(3) node[pos=0.5,sloped,above] {$K_2$};
\node (4) [circle, draw, fill=lightgray] at (6,2) {S4};
\draw(3)--(4) node[pos=0.5,sloped,above] {$K_1$};
\node (5) [circle, draw, fill=lightgray] at (8,2) {S5};
\draw(4)--(5) node[pos=0.5,sloped,above] {$K_2$};
\node (6) [circle, draw, fill=lightgray] at (10,2) {S6};
\draw(5)--(6) node[pos=0.5,sloped,above] {$K_1$};
\node (7) [circle, draw, fill=lightgray] at (12,2) {S7};
\draw(6)--(7) node[pos=0.5,sloped,above] {$K_2$};

\draw[->](5,1.5)--(5,0.5);
\end{tikzpicture}
\caption{Non similar decimation}
\end{figure}
\begin{eqnarray}
\text{tanh}(K_1')=\text{tanh}(K_1)\text{tanh}(K_2)
\end{eqnarray}
In this case a fixed point discussion is not possible as $K_2'$ does not exist. However, the free energy is the same in either case.
\section{Fibonacci chain Ising model}
In this section we consider a fibonacci series where two $b's$ are never adjacent $i.e.$ the nearest neighbour of 'b' is always 'a' (see Fig.(5)). We suggest  a method to generate the decimation procedure below. 
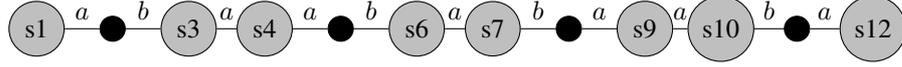
\begin{figure}[h]
\centering
\begin{tikzpicture}
\node (1) [circle, draw, fill=lightgray] at (0,2) {s1};
\node (2) [circle, draw, fill=black] at (1,2) {};
\draw(1)--(2) node[pos=0.5,sloped,above] {$a$};
\node (3) [circle, draw, fill=lightgray] at (2,2) {s3};
\draw(2)--(3) node[pos=0.5,sloped,above] {$b$};
\node (4) [circle, draw, fill=lightgray] at (3,2) {s4};
\draw(3)--(4) node[pos=0.5,sloped,above] {$a$};
\node (5) [circle, draw, fill=black] at (4,2) {};
\draw(4)--(5) node[pos=0.5,sloped,above] {$a$};
\node (6) [circle, draw, fill=lightgray] at (5,2) {s6};
\draw(5)--(6) node[pos=0.5,sloped,above] {$b$};
\node (7) [circle, draw, fill=lightgray] at (6,2) {s7};
\draw(6)--(7) node[pos=0.5,sloped,above] {$a$};
\node (8) [circle, draw, fill=black] at (7,2) {};
\draw(7)--(8) node[pos=0.5,sloped,above] {$b$};
\node (9) [circle, draw, fill=lightgray] at (8,2) {s9};
\draw(8)--(9) node[pos=0.5,sloped,above] {$a$};
\node (10) [circle, draw, fill=lightgray] at (9,2) {s10};
\draw(9)--(10) node[pos=0.5,sloped,above] {$a$};
\node (11) [circle, draw, fill=black] at (10,2) {};
\draw(10)--(11) node[pos=0.5,sloped,above] {$b$};
\node (12) [circle, draw, fill=lightgray] at (11,2) {s12};
\draw(11)--(12) node[pos=0.5,sloped,above] {$a$};

\end{tikzpicture}
\caption{Fibonacci Ising model }
\end{figure}
\subsection{Matrix method for generation}
It is well known that the Fiboanacci chain can be generated by setting up a number of rules for rabbit procreation, better known as Fibonacci Rabbits\cite{4}. In the present case we generate it by the mathematical operation shown below. Note that N$^{(0)}$ denotes a vector AB. N$^{(i)}$ denotes the resulting vector after N$^{(0)}$ has been operated 'i' times by M, a matrix operator to generate the Fibonacci Ising chain. 
\begin{eqnarray}
M=
  \left[ {\begin{array}{cc}
   1 & 1 \\
   1 & 0 \\
  \end{array} }\right], \quad
N^{(0)}=
  \left[ {\begin{array}{cc}
   N^{(0)_A} \\
   N^{(0)_B} \\
  \end{array} } \right]
=
  \left[ {\begin{array}{cc}
   A \\
   B \\
  \end{array} } \right] \colon AB
\end{eqnarray}
\begin{eqnarray}
MN^{(0)}=
  \left[ {\begin{array}{cc}
   1 & 1 \\
   1 & 0 \\
  \end{array} }\right]
  \left[ {\begin{array}{cc}
   A \\
   B \\
  \end{array} } \right]
=
  \left[ {\begin{array}{cc}
   A+B \\
   A \\
  \end{array} } \right] \colon ABA = N^{(1)}
\end{eqnarray}
\begin{eqnarray}
MN^{(1)}=
  \left[ {\begin{array}{cc}
   1 & 1 \\
   1 & 0 \\
  \end{array} }\right]
  \left[ {\begin{array}{cc}
   A+B \\
   A \\
  \end{array} } \right]
=
  \left[ {\begin{array}{cc}
   A+B+A \\
   A+B \\
  \end{array} } \right] 
=
  \left[ {\begin{array}{cc}
 N^{(2)_A} \\
   N^{(2)_B} \\
 \end{array} } \right] 
=N^{(2)}\colon ABAAB
\end{eqnarray}
\begin{eqnarray}
MN^{(2)}=
  \left[ {\begin{array}{cc}
   1 & 1 \\
   1 & 0 \\
  \end{array} }\right]
  \left[ {\begin{array}{cc}
   A+B+A \\
   A+B \\
  \end{array} } \right]
=
  \left[ {\begin{array}{cc}
   A+B+A+A+B \\
   A+B+A \\
  \end{array} } \right] 
=
  \left[ {\begin{array}{cc}
 N^{(3)_A} \\
   N^{(3)_B} \\
 \end{array} } \right] \\ 
=N^{(3)}\colon ABAABABA\nonumber
\end{eqnarray} and so on. 
\par For general iteration,
\begin{eqnarray}
N_A^{(n+1)}&=&M_{11}N_A^{(n)}+M_{21}N_B^{(n)}\quad M_{11}=M_{21}=M_{12}=1\\
N_B^{(n+1)}&=&M_{12}N_A^{(n)}+M_{22}N_B^{(n)}\quad\quad M_{22}=0
\end{eqnarray}
as $n\longrightarrow\infty$, the ratio $r=\text{lim}_{n\longrightarrow\infty} N^{(n)_A}/N^{(n)_B} \,\longrightarrow (\sqrt{5}+1)/2$. Here $N_A, N_B$ are length scales of bond A and B respectively.

\subsection{Decimation method}
\par Consider now the Ising Chain shown in Fig.(6): 
\newline
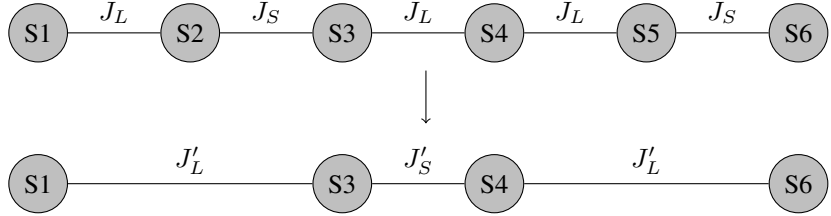
\begin{figure}[h]
\centering
\begin{tikzpicture}
\node (1) [circle, draw, fill=lightgray] at (0,2) {S1};
\node (2) [circle, draw, fill=lightgray] at (2,2) {S2};
\draw(1)--(2) node[pos=0.5,sloped,above] {$J_L$};
\node (3) [circle, draw, fill=lightgray] at (4,2) {S3};
\draw(2)--(3) node[pos=0.5,sloped,above] {$J_S$};
\node (4) [circle, draw, fill=lightgray] at (6,2) {S4};
\draw(3)--(4) node[pos=0.5,sloped,above] {$J_L$};
\node (5) [circle, draw, fill=lightgray] at (8,2) {S5};
\draw(4)--(5) node[pos=0.5,sloped,above] {$J_L$};
\node (6) [circle, draw, fill=lightgray] at (10,2) {S6};
\draw(5)--(6) node[pos=0.5,sloped,above] {$J_S$};
\node (7) [circle, draw, fill=lightgray] at (0,0) {S1};
\node (8) [circle, draw, fill=lightgray] at (4,0) {S3};
\draw(7)--(8) node[pos=0.5,sloped,above] {$J_L'$};
\node (9) [circle, draw, fill=lightgray] at (6,0) {S4};
\draw(8)--(9) node[pos=0.5,sloped,above] {$J_S'$};
\node (10) [circle, draw, fill=lightgray] at (10,0) {S6};
\draw(9)--(10) node[pos=0.5,sloped,above] {$J_L'$};

\draw [->] (5.1,1.5)--(5.1,0.8);
\end{tikzpicture}
\caption{Self-similar blocking for Fibonacci chain}
\end{figure}
\par Let $K_i= J_i/kT $
\begin{eqnarray}
H_N(K)&=&-\sum_{i=1}^{N-1}K_iS_iS_{i+1}\\
e^{K_L'S_1S_3}*e^{g_1}&=&\sum_{S_2=+1}^{-1}e^{K_LS_1S_2}*e^{K_SS_2S_3}
\end{eqnarray}
As in the previous sections, let $S_1=S_3=+1$ we obtain,
\begin{eqnarray}
e^{K_L'}*e^{2g_1}&=&e^{K_L+K_S}+e^{-(K_L+K_S)}
\end{eqnarray}
and $S_1=-S_3=+1$ yields,
\begin{eqnarray}
e^{-K_L'}*e^{2g_1}=e^{K_L-K_S}+e^{-(K_L-K_S)}
\end{eqnarray}
Using Eqs.(50) and (51),  and $cosh(x)=cosh(-x)$ leads to,
\begin{eqnarray}
tanh(K_L')&=&tanh(K_L)tanh(K_S)\\
\text{and }g_1&=&\frac{1}{2}K_L'+\frac{1}{2}ln(2cosh(K_L-K_S))
\end{eqnarray}
For the ordered case ($K_L = K_S$) the tansformation reduces to $tanh(K')=tanh^2(K)$ and $g_1$ is given by the same expression as Eq.(18). Further, $K_S'=K_L$ and $g_2=0$. Hence, all the above equations are consistent with ordered case.  The fixed points in this case are,$\{K_1^*,K_2^*\}=\{0,0\}$ or $\{\infty,\infty\}$.\par We designate bond lengths, distance between two neighbouring sites in the lattice before and after first decimation process, for the two types of bonds by two sets of variables. Here, the new lengths are L$'$ for K$_1'$ and S$'$ for K$_2'$ corresonding to the old lengths, L for K$_1$ and S for K$_2$. Self similarity is preserved if the new lengths follow the following relation.
\begin{eqnarray}
\frac{L'}{S'}=\frac{L}{S}
\end{eqnarray}
But $L'=L+S$ and $S'=L$. Thus,
\begin{eqnarray}
1+\frac{S}{L} &=&\frac{L}{S}=x\\
\ 1+\frac{1}{x}&=&x
\end{eqnarray}
which leads to $x=(\sqrt{5}+1)/2$, the golden ratio.
\section{Conclusion}
This decimation approach is perhaps the simplest version of RNG. Its extension to higher dimension however gets tricky. The solution to this problem uses the Migdal Kadanoff transformation\cite{5},\cite{6}. Interestingly, this decimation procedure inspired similar work in quantum systems. We hope to describe this quantum version introduced by Bhat, Singh and Subbarao\cite{7} in the future.
\section{Acknowledgment}
One of us (VAS) acknowledges support from the Raja Ramanna Fellowship by the DAE. 


\begin{thebibliography}{100}
\bibitem{1}Wilson K.G., Phys. Rev. B \textbf{4}, 1374, 1384 (1971).
\bibitem{2}Nauenberg M., J. Math. Phys. \textbf{16}, 703 (1975).
\bibitem{3}Wilson K.G., Rev. Mod. Phys. \textbf{47} , 773 (1975). 
\bibitem{4}Ball K. M., Strange Curves, Counting Rabbits, and Other Mathematical Explorations, Princeton University Press, \text{8}, (2003)
\bibitem{5}Kadanoff L.P., Ann. Phys. (N.Y.) \textbf{100}, 359 (1976). 
\bibitem{6}Migdal A.A., Soy. Phys. JETP \textbf{42}, 743 (1976).
\bibitem{7}Bhat G.R. Singh V.A. and Subbarao K., J. Phys. C, Solid State Physics, \textbf{17},  5569, (1984).
\end{thebibliography}
\end{document}